\documentclass[12pt,preprint]{aastex}

\shorttitle{Fe XI lines in SERTS active region spectrum}
\shortauthors{F. P. Keenan et al.}
\begin{document}

\title{\ion{Fe}{11} emission lines in a high resolution extreme ultraviolet
active region spectrum obtained by SERTS}

\author{F. P. Keenan\altaffilmark{1}, K. M. Aggarwal\altaffilmark{1},
R. S. I. Ryans\altaffilmark{1}, R. O. Milligan\altaffilmark{1,2},
D. S. Bloomfield\altaffilmark{1}, J. W. Brosius\altaffilmark{2,3}, 
J. M. Davila\altaffilmark{2} and R. J. Thomas\altaffilmark{2}}

\email{F.Keenan@qub.ac.uk}

\altaffiltext{1}{Department of Pure and Applied Physics, Queen's University Belfast,
Belfast, BT7 1NN, Northern Ireland, U.K.}

\altaffiltext{2}{Laboratory for Astronomy and Solar Physics, Code 682,
NASA's Goddard Space Flight Center, Greenbelt, MD 20771}

\altaffiltext{3}{Department of Physics, The Catholic University of
America, Washington, DC 20064}



\begin{abstract}

New calculations of radiative rates and electron impact excitation cross sections for 
\ion{Fe}{11} are used to derive emission line intensity ratios involving
3{\em s}$^{2}$3{\em p}$^{4}$--3{\em s}$^{2}$3{\em p}$^{3}$3{\em d} transitions
in the 180--223 \AA\ wavelength range. These ratios are subsequently compared
with observations of a solar active region, obtained during the 1995 flight
of the {\em Solar EUV Research Telescope and Spectrograph} (SERTS). The version of
SERTS flown in 1995 incorporated a multilayer grating that enhanced the instrumental 
sensitivity for features
in the $\sim$170--225 \AA\ wavelength range,
observed in second-order between 340 and 450 \AA. This enhancement led
to the detection of many emission lines not seen on previous SERTS flights, which were
measured with the highest spectral resolution (0.03 \AA) ever achieved 
for spatially resolved active region spectra in this wavelength range.
However, even at this high spectral resolution, several of the \ion{Fe}{11}
lines are found to be blended, although the sources of the blends are
identified in the majority of cases.
The 
most useful \ion{Fe}{11} electron density diagnostic line intensity ratio
is I(184.80 \AA)/I(188.21 \AA). This ratio involves lines close
in wavelength and free from blends, and which 
varies by a factor of 11.7
between
N$_{e}$ = 10$^{9}$ and 10$^{11}$ cm$^{-3}$, yet shows little temperature
sensitivity. An unknown line in the SERTS spectrum at 189.00 \AA\ is found to
be due to \ion{Fe}{11}, the first time (to our knowledge) this feature
has been identified in the solar spectrum. Similarly, there are new
identifications of the \ion{Fe}{11} 192.88,
198.56 and 202.42 \AA\ features, although the latter two are
blended with \ion{S}{8}/\ion{Fe}{12} and \ion{Fe}{13}, respectively.

\end{abstract}

\keywords{Atomic data --
	Sun: active region --
	Ultraviolet: spectra}


\section{Introduction}

Emission features arising from transitions in \ion{Fe}{11}
are routinely detected in solar extreme ultraviolet spectra
(see, for example,
Dere 1978; Thomas \& Neupert 1994).
The diagnostic potential of these lines for determining the
electron density in the emitting plasma was first demonstrated 
by Kastner \& Mason (1978). Since then, several authors have produced
theoretical line ratios for \ion{Fe}{11}
applicable to solar spectra (see Bhatia, Doschek, \& Eissner 2002
and references
therein).

In this paper we present theoretical \ion{Fe}{11} line ratios for a range of
electron temperatures and densities, generated using the most
recent calculations of radiative rates and R-matrix electron impact
excitation cross sections. We subsequently
compare these line ratios with active region observations
obtained with the {\em
Solar EUV Research Telescope and Spectrograph} (SERTS) 
sounding rocket experiment. 
Specifically, we employ the SERTS dataset obtained during the 1995 flight,
when the instrument incorporated a multilayer-coated toroidal diffraction
grating that enhanced its sensitivity for features in the 
$\sim$170--225 \AA\ wavelength range, observed in second-order
between 340 and 450 \AA.  
This led to the detection of many emission lines not seen on previous
SERTS flights (Thomas \& Neupert 1994; Brosius et al.
1996), and provided the highest spectral resolution (0.03 \AA) ever achieved
for spatially resolved active region spectra in this wavelength range
(Brosius, Davila, \& Thomas 1998). Hence this SERTS dataset allows us to
undertake the most comprehensive 
analysis to date of the solar \ion{Fe}{11} spectrum in the
170--225 \AA\ wavelength region.
  
\section{Observational Data}

The solar spectrum analysed in the present paper is that
of active region
NOAA 7870, recorded on Eastman Kodak 101--07 emulsion
by SERTS during a rocket flight on 1995 May 15 at 1800 UT
(Brosius et al. 1998). SERTS
first flew in 1989 (Neupert et al. 1992; Thomas \& Neupert 
1994), and carried a standard gold-coated toroidal diffraction
grating. It observed hundreds of first-order emission lines 
in the 235--450 \AA\ wavelength range, as well as dozens of 
features spanning 170--225 \AA, which appeared in second-order
between 340 and 450 \AA.
The version of SERTS flown in 1991 and 1993 (Brosius et al. 1996) carried
a multilayer-coated diffraction grating that enhanced the 
instrumental efficiency in the first-order wavelength range.
However, as noted in \S\ 1, the version flown in 
1995 incorporated a multilayer grating that enhanced the
instrumental sensitivity for second-order features.
The SERTS 1995 active region spectrum therefore  
provides the best observations for investigating
solar \ion{Fe}{11} emission lines in the 170--225 \AA\
wavelength region.
Further details of the observations, and the wavelength and
absolute flux calibration procedures
employed in the data reduction,
may be found in Brosius et al. (1998). 

We have searched for \ion{Fe}{11} 
emission lines in the SERTS spectrum, using the 
identifications of Jup\'{e}n, Isler, \& Tr\"{a}bert
(1993) and the NIST database\footnote{http://physics.nist.gov/PhysRefData/},
as well as previous solar detections where available (for example,
those by Behring, Cohen, \& Feldman 1972; Behring et al. 1976). In Table 1
we list the \ion{Fe}{11} 
transitions
found in the spectrum,
along with the measured wavelengths.
We also note possible blending species or alternative
identifications, as suggested by Brosius et al. (1998) in their
original line list for the active region spectrum,

Intensities and
line widths (FWHM) of the \ion{Fe}{11} features are given in 
Table 2, along with the associated 1$\sigma$ errors. These were
determined by using the
spectrum synthesis package {\sc dipso} (Howarth, Murray, \& Mills
1994) to fit Gaussian profiles to
the observations. Uncertainties in these measurements
have been determined using methods
discussed in detail by Thomas \& Neupert (1994).
In Figures 1--4 we
plot portions of the SERTS spectrum containing the
\ion{Fe}{11}
features, to show the quality of the observational data.
We note that each SERTS spectrum exhibits
a background level due to film fog, scattered light and actual
solar continuum. The background was calculated using methods
detailed in Thomas \& Neupert (1994) and Brosius et al. (1998),
and then subtracted from the initial spectrum, leaving
only an emission line spectrum (with noise) on a zero base level.
It is this zero base level which is shown in Figures 1--4.
We note that some of the measured \ion{Fe}{11} emission
lines, such as the 184.80 \AA\ transition (Figure 1), have
line intensities comparable to the noise fluctuations.
In these instances the reality of the line was confirmed
by a visual inspection of the original SERTS film.

\section{Theoretical Line Ratios} 

The model ion for \ion{Fe}{11} consisted of the 24
energetically lowest LS states, namely
3{\em s}$^{2}$3{\em p}$^{4}$ $^{3}$P, $^{1}$D, $^{1}$S; 
3{\em s}3{\em p}$^{5}$ $^{3}$P, $^{1}$P; 
3{\em s}$^{2}$3{\em p}$^{3}$($^{4}$S)3{\em d} $^{5}$D;
3{\em s}$^{2}$3{\em p}$^{3}$($^{2}$D)3{\em d} $^{3}$D,
$^{3}$F, $^{1}$S, $^{3}$G, $^{1}$G; 
3{\em s}$^{2}$3{\em p}$^{3}$($^{2}$P)3{\em d} $^{1}$D, $^{3}$D, $^{3}$P, $^{3}$F,
$^{1}$F; 
3{\em s}$^{2}$3{\em p}$^{3}$($^{2}$D)3{\em d} $^{3}$S, $^{3}$P, $^{1}$P;
3{\em s}$^{2}$3{\em p}$^{3}$($^{4}$S)3{\em d} $^{3}$D;
3{\em s}$^{2}$3{\em p}$^{3}$($^{2}$D)3{\em d} $^{1}$D, $^{1}$F, $^{1}$P;
3{\em p}$^{6}$ $^{1}$S, yielding a total of 48 fine-structure levels.
Experimental energy levels, which are only available for a relatively
small number (20) of \ion{Fe}{11} states, were obtained from
Shirai et al. (1990) and Jup\'{e}n et al. (1993).
For the remaining values the theoretical results of Aggarwal \& Keenan (2003a)
were adopted. Test calculations including higher-lying 
3{\em s}$^{2}$3{\em p}$^{3}$4{\em l} levels were found to
have a negligible effect on the theoretical line ratios considered
in this paper.

The electron impact excitation cross sections adopted in the present paper
are the recent R-matrix calculations of Aggarwal \& Keenan (2003b), 
while Einstein A-coefficients 
for allowed and intercombination lines were obtained from Aggarwal \& Keenan
(2003a). These A-values are similar to those of Deb \& Tayal (1998) and
Bhatia \& Doschek (1996), apart from transitions involving 
levels 39 (3{\em s}$^{2}$3{\em p}$^{3}$($^{2}$D)3{\em d} $^{3}$P$_{1}$) 
and 41 (3{\em s}$^{2}$3{\em p}$^{3}$($^{2}$D)3{\em d} $^{1}$P$_{1}$),
where there are large discrepancies. It appears that these levels have
been interchanged in the previous two calculations, as agreement
is restored if Aggarwal \& Keenan
(2003a) reclassify the A-values
relating to level 39 as belonging to level 41, and vice-versa.
However we note that
Aggarwal \& Keenan (2003a) performed several test calculations,
involving configuration interaction with different orbitals and configurations, and
in each instance obtained the same energy 
level ordering. They are
hence confident of the
ordering of the levels of their calculations.

Radiative rates for forbidden transitions have also been
taken from the work of Aggarwal \& Keenan (2003a), although the data were
not included in the published paper for conciseness. They
are available from the authors on request. However we note that
the results are very similar to those calculated by others, such
as Bhatia et al. (2002). Proton impact excitation is only important for
transitions within the 3{\em s}$^{2}$3{\em p}$^{4}$ $^{3}$P ground term, 
and in the present analysis we have employed the calculations of
Landman (1980).

Using the above atomic data, in conjunction
with a recently updated version of the statistical equilibrium
code of Dufton (1977), relative \ion{Fe}{11}
level populations and hence emission line
strengths were calculated as a function of electron temperature
(T$_{e}$) and density (N$_{e}$).
Details of the procedures involved and approximations made
may be found in
Dufton (1977) and Dufton et al.
(1978).

In Figures 5--9 we plot the theoretical
emission line ratios

R$_{1}$ = I(180.38 \AA)/I(188.21 \AA),

R$_{2}$ = I(181.13 \AA)/I(188.21 \AA),

R$_{3}$ = I(182.17 \AA)/I(188.21 \AA),

R$_{4}$ = I(184.80 \AA)/I(188.21 \AA),

R$_{5}$ = I(188.30 \AA)/I(188.21 \AA),

R$_{6}$ = I(189.00 \AA)/I(188.21 \AA),

R$_{7}$ = I(189.19 \AA)/I(188.21 \AA),

R$_{8}$ = I(202.42 \AA)/I(188.21 \AA),

and

R$_{9}$ = I(223.00 \AA)/I(188.21 \AA),

as a function of 
electron density at the temperature
of maximum \ion{Fe}{11} fractional abundance in ionization
equilibrium, T$_{e}$ = T$_{max}$ = 
10$^{6.1}$ K, plus $\pm$0.2 dex
about this value, where the fractional abundance has fallen to
N(\ion{Fe}{11}/N(Fe) $\leq$ 0.06 (Mazzotta et al. 1998).

We note that the ratios

R$_{10}$ = I(189.72 \AA)/I(188.21 \AA),

R$_{11}$ = I(192.88 \AA)/I(188.21 \AA),

R$_{12}$ = I(193.51 \AA)/I(188.21 \AA),

and

R$_{13}$ = I(198.56 \AA)/I(188.21 \AA),

have the same temperature and density dependence
as R$_{5}$ or R$_{7}$,
owing to common upper
levels, but with

R$_{10}$ = 0.875 $\times$ R$_{7}$, 

R$_{11}$ = 0.337 $\times$ R$_{5}$,

R$_{12}$ = 0.0929 $\times$ R$_{5}$,

and 

R$_{13}$ = 0.438 $\times$ R$_{7}$.

Similarly, the branching
ratio

R$_{14}$ = I(192.81 \AA)/I(188.21 \AA),

is predicted to have the constant value R$_{14}$ = 0.190,
owing to common upper levels.

The transitions corresponding to the
wavelengths listed above are given in Table 1.
Given errors in the adopted atomic data of typically
$\pm$10\%\ (see the references above), we would expect the
theoretical ratios to be accurate to better than $\pm$20\%.

The ratios in Figures 5--9 are given relative to the 188.21 \AA\
transition, as this feature is the cleanest and 
most reliably detected
\ion{Fe}{11}
emission line in the SERTS spectrum (see Figures 1--4).
We have confirmed this via a search of line lists, 
such as the Atomic Line List of van 
Hoof\footnote{http://star.pst.qub.ac.uk/$\sim$pvh/},
and also by generating
a synthetic active region spectrum using the latest version
(4.2) of the {\sc chianti} database (Dere et al. 1997;
Young et al. 2003). No transitions in first or
second-order with intensities $>$ 3\%\ that of the 188.21 \AA\
line were found.
However, we note that theoretical results
involving any line pair
are available from one of the authors
(F.Keenan@qub.ac.uk)
on request.

An inspection of the figures reveals that several of the
line ratios, in particular R$_{3}$ and R$_{4}$, are
very sensitive to variations in the electron density. However
most of the ratios are relatively insensitive to the adopted
electron temperature.  The strong N$_{e}$--dependence of ratios
such as R$_{3}$ and R$_{4}$, combined 
with their temperature insensitivity, 
make them potentially useful
density diagnostics
for the \ion{Fe}{11} emitting region
of a plasma.

We note that the current line ratios can differ by up to
typically 20--30\%\ with other recent calculations. For example,
at T$_{e}$ = 10$^{6.1}$ K and N$_{e}$ =
10$^{8}$ cm$^{-3}$, we calculate R$_{1}$ = 1.8 and R$_{3}$ = 0.18,
compared to R$_{1}$ = 2.2 and R$_{3}$ = 0.26 from the latest 
version of {\sc chianti}. 
These discrepancies are due primarily to the adoption
of the Aggarwal \& Keenan (2003b) excitation rates in the present
analysis, as opposed to the results of
Bhatia \& Doschek (1996) and Gupta \& Tayal (1999a,b) adopted by
previous authors (see Aggarwal \& Keenan 2003b for more details).

\section{Results and Discussion}

In Table 3 we summarise the observed \ion{Fe}{11} emission line
intensity ratios, along with the
associated 1$\sigma$ errors. Also shown in the table
are the theoretical results from Figures 5--9 at 
the temperature of maximum fractional abundance
in ionization equilibrium for \ion{Fe}{11}, T$_{max}$ = 10$^{6.1}$ K
(Mazzotta et al. 1998), and an electron density
of N$_{e}$ = 10$^{9.4}$ cm$^{-3}$. This density was derived for the
SERTS active region from emission 
line ratios in \ion{Fe}{10}, \ion{Fe}{13} and \ion{Fe}{14} 
(Brosius et al. 1998), which have similar values of T$_{max}$ to
\ion{Fe}{11} (10$^{6}$--10$^{6.3}$ K).
Hence the \ion{Fe}{10}, \ion{Fe}{13} and \ion{Fe}{14} densities 
should reflect that of the \ion{Fe}{11}
emitting plasma in the active region.
We note that \ion{Fe}{12} line ratios in the active region
imply significantly higher values of density (N$_{e}$ $\simeq$ 10$^{10}$ cm$^{-3}$),
but these have not been adopted in the present work as they are inconsistent
with the other Fe ions, and may indicate a problem with the
\ion{Fe}{12} diagnostics. However, changing the theoretical line ratios
in Table 3 to values for N$_{e}$ $\simeq$ 10$^{10}$ cm$^{-3}$ would not
significantly alter the discussion below. 
For R$_{14}$, which is predicted to 
be N$_{e}$--insensitive, the theoretical result in Table 3 
is the value listed in \S\ 3.
Error bars for all the theoretical ratios are based on the
estimated $\pm$20\%\ accuracy of the calculations (see \S\ 3).

An inspection of Table 3 reveals excellent agreement between theory and
observation for the R$_{4}$ and R$_{14}$ ratios, indicating that the
184.80 and 192.81 \AA\ lines are well detected and free from blends.
More importantly, these results imply that R$_{4}$ may be employed with
confidence as an N$_{e}$--diagnostic. It should be particularly useful over 
the N$_{e}$ = 10$^{9}$--10$^{11}$ cm$^{-3}$ interval, where it varies
by a factor of 11.7 and yet shows little temperature sensitivity (see
Figure 7).

For the ratios R$_{2}$, R$_{3}$ and R$_{10}$, agreement between theory and
observation is less satisfactory, but the calculated and experimental 
measurements do overlap within the uncertainties. Hence we can state that
the 181.13, 182.17 and 189.72 \AA\ emission lines appear to be free
from blends, especially as an inspection of line lists and the synthetic
active region spectrum from {\sc chianti}
reveals no likely candidates in first or
second-order.
From Figure 6, the R$_{3}$ ratio should provide a reasonably
good N$_{e}$--diagnostic, 
varying by a factor of 2.5 between 
N$_{e}$ = 10$^{9}$ and 10$^{11}$ cm$^{-3}$. However 
on the
basis of the present analysis it is advised that more weight should be
given to results from R$_{4}$.

In the cases of R$_{1}$ and R$_{12}$, the measured line ratios are much
larger than the theoretical values, confirming that the 180.38 and 193.51 \AA\
lines are badly blended. Brosius et al. (1998) note that the 180.38 \AA\
line is blended with an \ion{Fe}{16} transition observed at 360.76 \AA\ 
in first-order. However these authors list the 193.51 \AA\ feature as
being due entirely to \ion{Fe}{12}. This is effectively the case, as \ion{Fe}{11}
contributes less than 2\%\ to the total 193.51 \AA\ line intensity.

Dere et al. (1997) list the 188.30 \AA\ line as being due to 
the 
3{\em s}$^{2}$3{\em p}$^{4}$ $^{3}$P$_{2}$--3{\em s}$^{2}$3{\em 
p}$^{3}$($^{2}$D)3{\em d} $^{1}$P$_{1}$ transition (equivalent to 
3{\em s}$^{2}$3{\em p}$^{4}$ $^{3}$P$_{2}$--3{\em s}$^{2}$3{\em 
p}$^{3}$($^{2}$D)3{\em d} $^{3}$P$_{1}$ in our notation, as we interchange
the $^{3}$P$_{1}$ and $^{1}$P$_{1}$ levels; see \S\ 3). They
preferred this to the 
3{\em s}$^{2}$3{\em p}$^{4}$ $^{3}$P$_{2}$--3{\em s}$^{2}$3{\em 
p}$^{3}$($^{2}$D)3{\em d} $^{3}$S$_{1}$ identification of 
Jup\'{e}n et al. (1993), as the latter was predicted to be weak.
However we calculate that the intensity of the $^{3}$P$_{2}$--$^{3}$S$_{1}$
line should be 21\%\ that of the 188.21 \AA\ feature, slightly larger
than the $^{3}$P$_{2}$--$^{3}$P$_{1}$ intensity which is predicted to
be 19\%\ of I(188.21 \AA). Hence we believe that the 188.30 \AA\ feature
is due to the $^{3}$P$_{2}$--$^{3}$S$_{1}$ transition, as
classified by Jup\'{e}n et al., although the disagreement between theory and
observation for R$_{5}$ implies that the line is blended, as previously
noted by
Brosius et al. (1998). There are no obvious blending species, although Brosius
et al. point out that Kelly (1987) lists an Si line from an undetermined ionization stage
at 188.3 \AA. Also, Dere (1978) lists a line at 376.61 \AA\ 
which would be the second-order detection of a feature at 
188.31 \AA. However Thomas \& Neupert (1994) identify this
as a \ion{Mg}{5} transition, which is very weak in the 1989 SERTS active region
dataset, and hence is unlikely to be responsible for the blend. 
This is confirmed by the {\sc chianti} synthetic spectrum, which indicates that
the \ion{Mg}{5} line should have an intensity 
$<$ 10$^{-5}$ that of the
\ion{Fe}{16} 360.76 \AA\ feature, observed by SERTS at 180.38 \AA\
with I = 3440 erg cm$^{-2}$ s$^{-1}$ sr$^{-1}$.
Hence the predicted intensity of the \ion{Mg}{5} transition
is $<$ 0.01\%\ that of the 188.30 \AA\ feature in Table 2.
  
Brosius et al. (1998) have tentatively associated the 189.19 \AA\ line in the
SERTS active region spectrum as being due to \ion{Fe}{11}, although the
wavelength of the feature is higher than previously measured (189.13 \AA).
Kelly (1987) lists an \ion{Mn}{9} line at 189.16 \AA\ and a \ion{Ni}{15}
transition at 189.21 \AA. However, a comparison of the observed and theoretical
R$_{7}$ ratios from Table 3 indicates that blending is probably not
responsible for any wavelength discrepancy, as the measured ratio
is actually smaller than the calculated value. The SERTS wavelength measurement
should be accurate to $\pm$0.01 \AA\ or better 
(see Figure 3 and Brosius et al.), and as noted in \S\ 2 this is
the highest resolution solar active region spectrum obtained over the
170--225 \AA\ wavelength region.
Hence we are confident of the wavelength measurement for the \ion{Fe}{11}
line, and the discrepancy may be due to inaccuracies in previous
determinations from lower quality data. 

The good agreement between theory and observation for R$_{6}$ confirms that
the 189.00 \AA\ feature, previously unidentified in the SERTS spectrum,
is due to \ion{Fe}{11}. To our knowledge, this is the first time this \ion{Fe}{11}
transition has been identified in the solar spectrum.
Similarly, our results for R$_{11}$ indicate that the 
192.88 \AA\ feature must have a significant contribution from \ion{Fe}{11}, and is
not totally due to \ion{Ca}{17} as suggested by Brosius et al. (1998).
The observed R$_{11}$ ratio is somewhat larger than theory (although they agree
within the error bars), so that \ion{Ca}{17} may make a 30--50\%\
contribution to the line blend. 
Once again, to our knowledge this is the first time this
\ion{Fe}{11} line has been identified in the solar spectrum.

Brosius et al. (1998) list the 223.00 \AA\ line as a blend of \ion{Fe}{11}
and \ion{Ca}{17}. However our calculations for R$_{9}$ show that the
\ion{Fe}{11}
contribution is negligible, while the intensity of 
the \ion{Ca}{17} 223.00 \AA\ transition
is predicted to be less than 1\%\ that of the 192.88 \AA\ line (Dufton
et al. 1983), ruling out a \ion{Ca}{17}
identification. 
A more likely candidate for the 223.00 \AA\ feature is the 
2{\em s} $^{2}$S$_{1/2}$--2{\em p} $^{2}$P$_{3/2}$ resonance line
of \ion{Cr}{22}, suggested by both Dere (1978) and Dufton et al.
Support for this comes from the {\sc chianti} synthetic spectrum,
which predicts the 2{\em s} $^{2}$S$_{1/2}$--2{\em p} $^{2}$P$_{1/2}$
line of \ion{Cr}{22} at 279.74 \AA\ and an intensity ratio
I(279.74 \AA)/I(223.00 \AA) = 0.42. There is indeed an unidentified
feature in the SERTS spectrum at this wavelength, with I(279.74 \AA)/I(223.00 \AA)
= 0.55 $\pm$0.21. The presence of two features
at the predicted wavelengths of the \ion{Cr}{22} lines and with the
correct intensity ratio may be a coincidence, but this must be
considered unlikely. Hence we believe that the 223.00 and 279.74 \AA\ features
are the two components of the \ion{Cr}{22} doublet.
Furthermore, the detection of the resonance lines of a highly
ionized species such as \ion{Cr}{22} in the SERTS spectrum implies that
\ion{Ca}{17} probably does make a significant contribution 
to the 192.88 \AA\ feature (see above), as the \ion{Ca}{17}
transition is also a strong resonance line, and the solar abundance
of Ca is larger than that for Cr.

The 198.56 and 202.42 \AA\ features are listed as being due to
\ion{S}{8}/\ion{Fe}{12} and \ion{Fe}{13}, respectively,
by Brosius et al. (1998). However our calculations for R$_{13}$
and R$_{8}$ reveal that \ion{Fe}{11} makes significant ($\sim$50\%)
contributions to both the 198.56 and 202.42 \AA\ line fluxes. Once again, this is
the first time (to our knowledge) that these \ion{Fe}{11} 
transitions have been identified in the solar spectrum.

One concern of the present work is our failure to 
detect the
3{\em s}$^{2}$3{\em p}$^{4}$ $^{3}$P$_{2}$--3{\em s}$^{2}$3{\em 
p}$^{3}$($^{2}$D)3{\em d} $^{3}$P$_{1}$ transition at 184.70 \AA\
(Jup\'{e}n et al. 1993), although we have identified the
$^{3}$P$_{1}$--$^{3}$P$_{1}$ and 
$^{3}$P$_{0}$--$^{3}$P$_{1}$ lines at 189.19 and 189.72 \AA,
respectively. 
The 184.70 \AA\ transition is predicted to have an intensity 
about 20\%\ that of the 188.21 \AA\ feature. However an inspection
of the SERTS dataset in this wavelength region (Figure 1) reveals
no convincing detection.

Clearly, further research on the \ion{Fe}{11}
spectrum is required. In particular, we believe that 
high spectral resolution observations of magnetically-confined
tokamak plasmas could provide an enormous contribution to
understanding solar \ion{Fe}{11} emission. The physical conditions in such
plasmas are quite similar to those found in the solar transition region
and corona, but can be independently measured to a high
degree of accuracy, which in turn allows theoretical line strengths
to be reliably predicted. Also, emission lines from species other
than the element under consideration will generally not play an important
role in a tokamak spectrum, greatly reducing the amount of blending.
As result, tokamak observations   
have frequently been employed to
test astrophysical diagnostic calculations, and 
to help
in the identification of emission lines in solar spectra
(see, for example, Keenan et al. 2000, 2003).
Through a collaboration with the UKAEA Culham Laboratory, we have access to
an extensive range of tokamak spectral observations, including those from the
{\em Joint European Torus} (JET). In the future we therefore plan to search
the JET and other tokamak databases for suitable \ion{Fe}{11} spectra.

\acknowledgements

K.M.A. and R.S.I.R. acknowledge financial support from the 
EPSRC and PPARC 
Research Councils of the
United Kingdom. 
R.O.M. and D.S.B are grateful to the Department of Education and Learning
(Northern Ireland) 
for the award of studentships, while the latter also
acknowledges financial support from 
NASA's Goddard Space Flight Center. 
The SERTS rocket programme is
supported by RTOP grants from the Solar Physics Office   
of NASA's Space Physics Division.                        
JWB acknowledges additional NASA support under
grant NAG5--13321.                                       
F.P.K. is grateful to AWE Aldermaston for the award of a William Penney
Fellowship. The authors thank Peter van Hoof for the use of his
Atomic Line List.

\clearpage


\begin{figure}
\includegraphics[scale=0.7,angle=90]{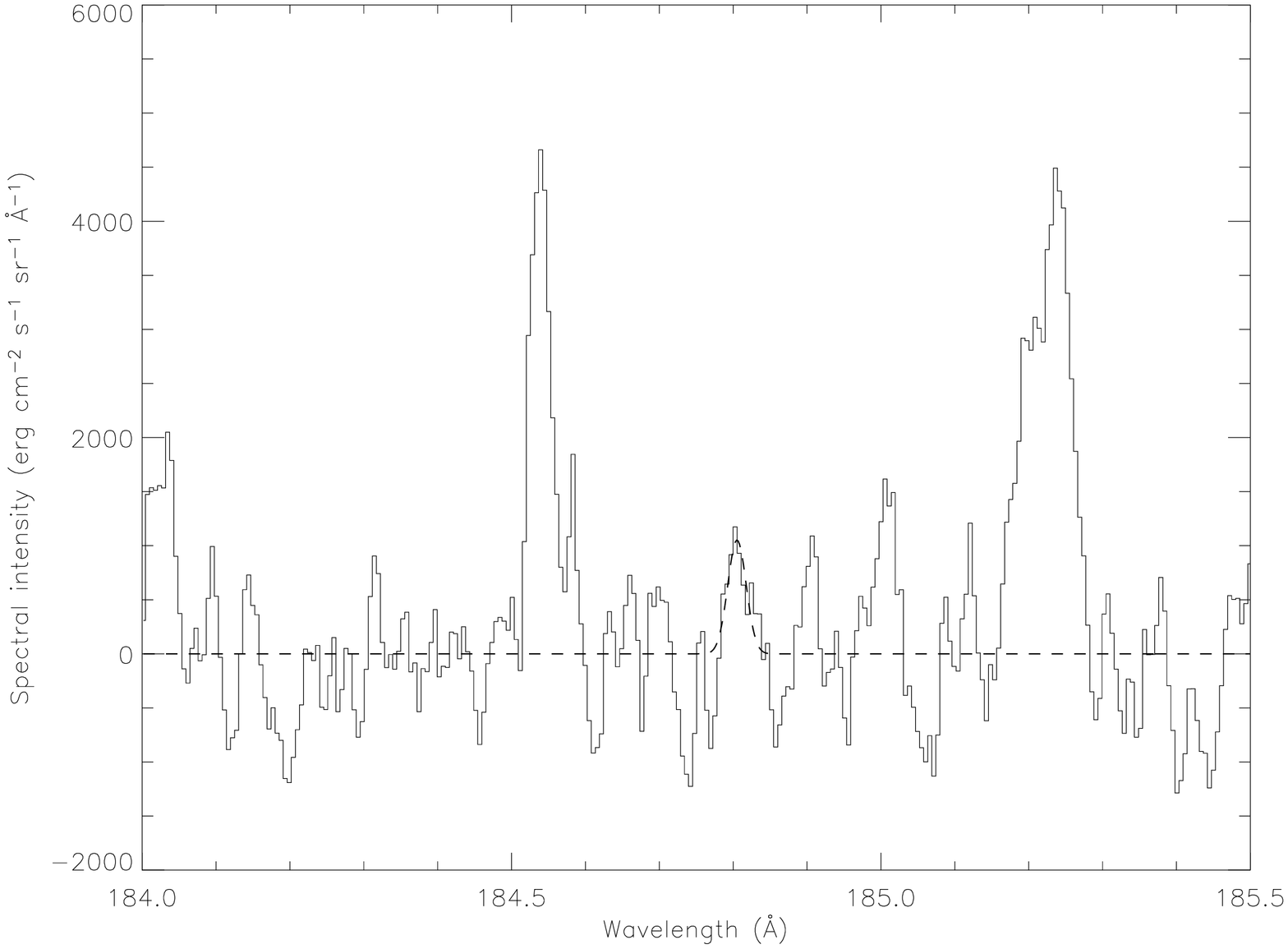}
\caption{
Plot of the SERTS 1995 active region spectrum in the
184.0--185.5 \AA\ wavelength range.
The profile fit to the \ion{Fe}{11} 184.80 \AA\ feature
is shown by a dashed line. Also clearly visible in the
figure are the \ion{Fe}{10} 184.53 \AA\ and \ion{Fe}{8}/\ion{Ni}{16}
185.22 \AA\
lines.
}
\end{figure}

\begin{figure}
\includegraphics[scale=0.7,angle=90]{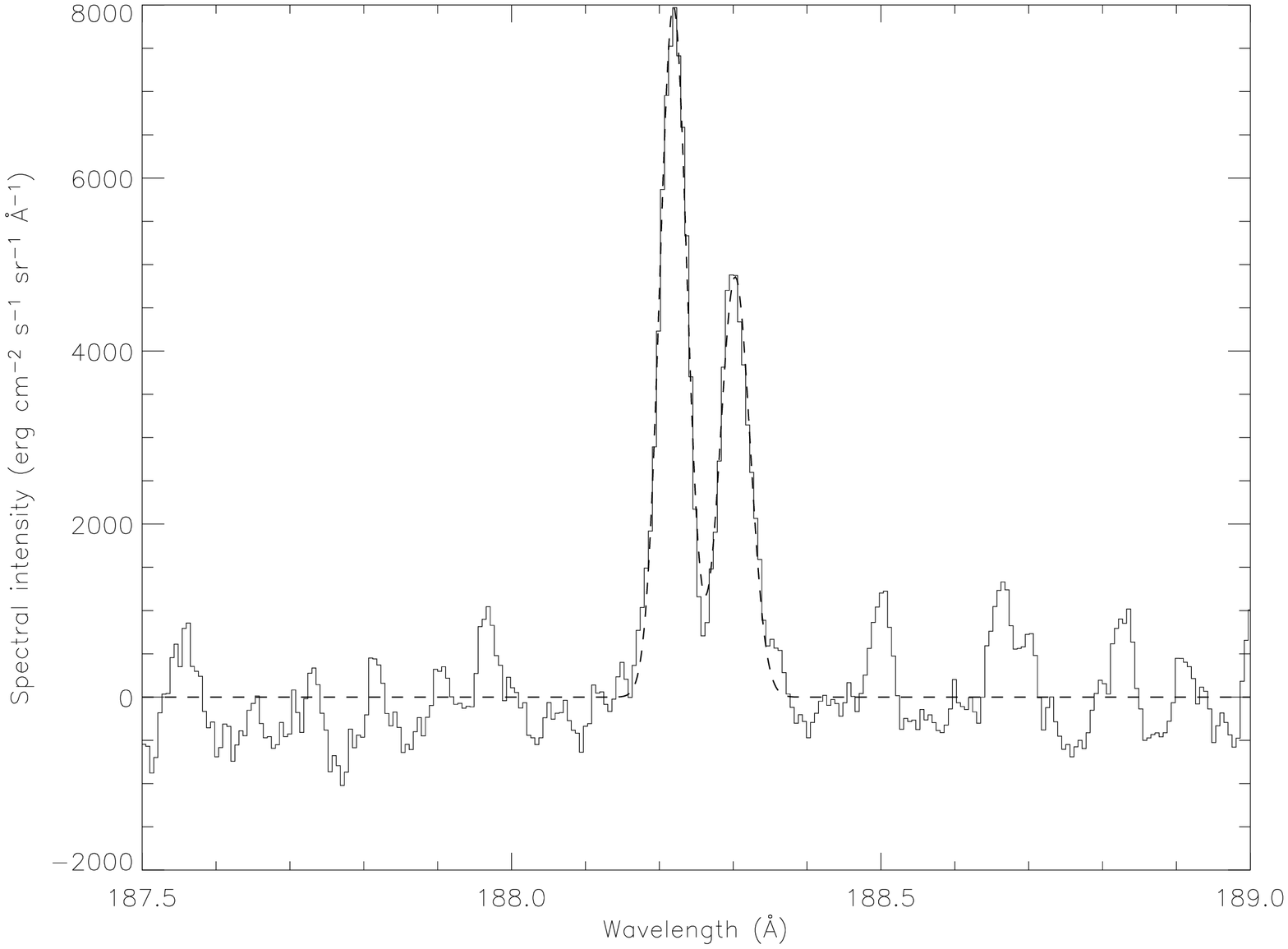}
\caption{
Plot of the SERTS 1995 active region spectrum in the
187.5--189.0 \AA\ wavelength range.
The profile fit to the \ion{Fe}{11} 188.21 and
188.30 \AA\ features
is shown by a dashed line. Also clearly visible in the
figure are the \ion{S}{11} 188.66 \AA\ and \ion{Ar}{11} 188.82 \AA\
lines.
}
\end{figure}

\begin{figure}
\includegraphics[scale=0.7,angle=90]{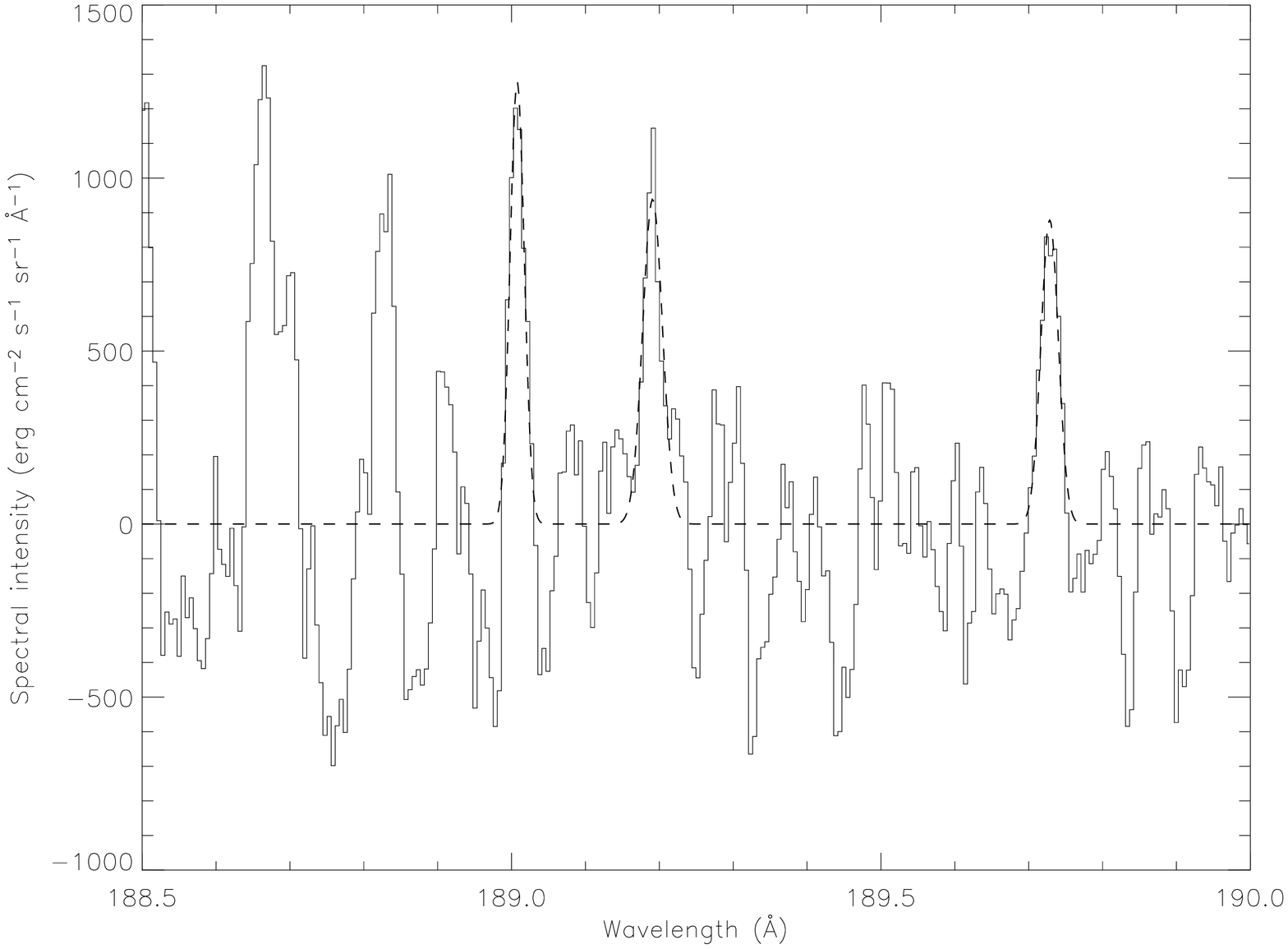}
\caption{
Plot of the SERTS 1995 active region spectrum in the
188.5--190.0 \AA\ wavelength range.
The profile fit to the \ion{Fe}{11} 189.00, 189.19 and
189.72 \AA\ features
is shown by a dashed line. Also clearly visible in the
figure are the \ion{S}{11} 188.66 \AA\ and \ion{Ar}{11} 188.82 \AA\
lines.
}
\end{figure}

\begin{figure}
\includegraphics[scale=0.7,angle=90]{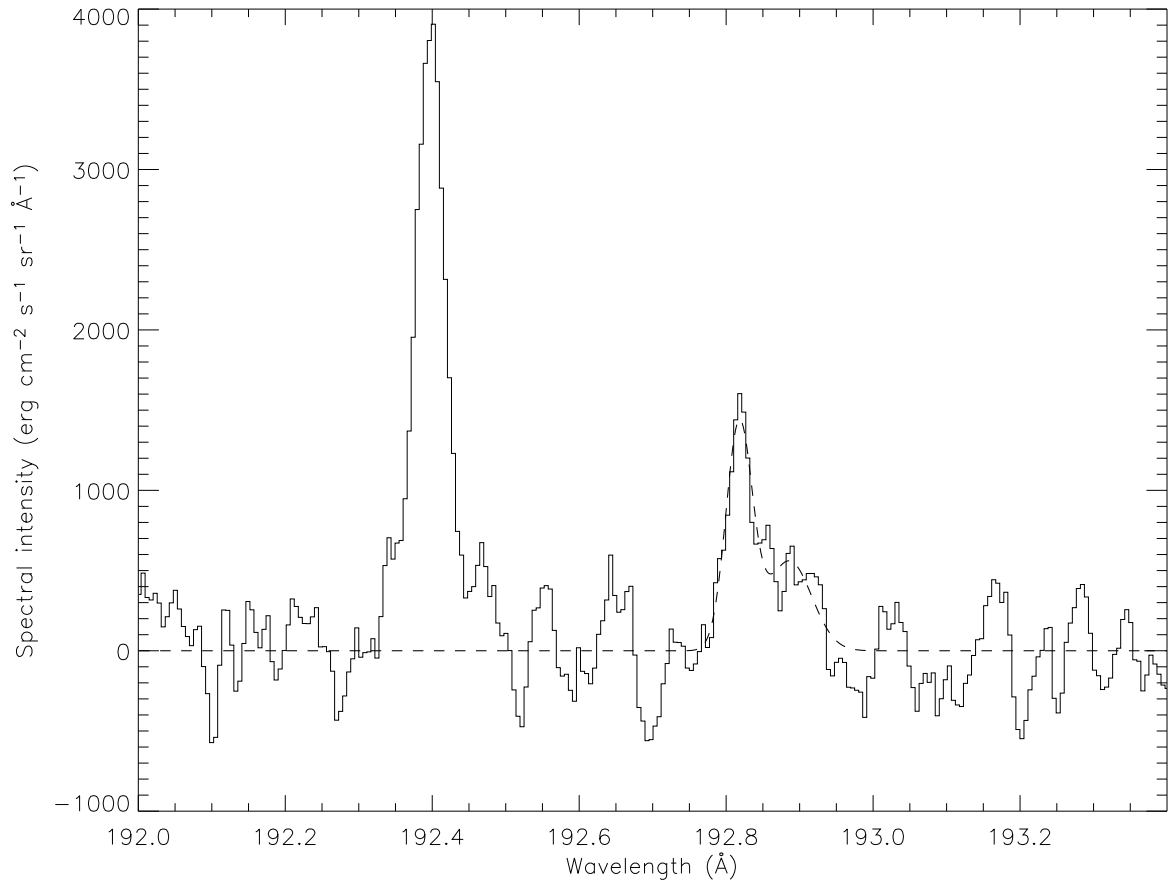}
\caption{
Plot of the SERTS 1995 active region spectrum in the
192.0--193.4 \AA\ wavelength range.
The profile fit to the \ion{Fe}{11} 192.81 and 192.88 
\AA\ features
is shown by a dashed line. Also clearly visible in the
figure is the \ion{Fe}{12} 192.39 \AA\ line.
}
\end{figure}

\begin{figure}
\includegraphics[scale=0.3,angle=90]{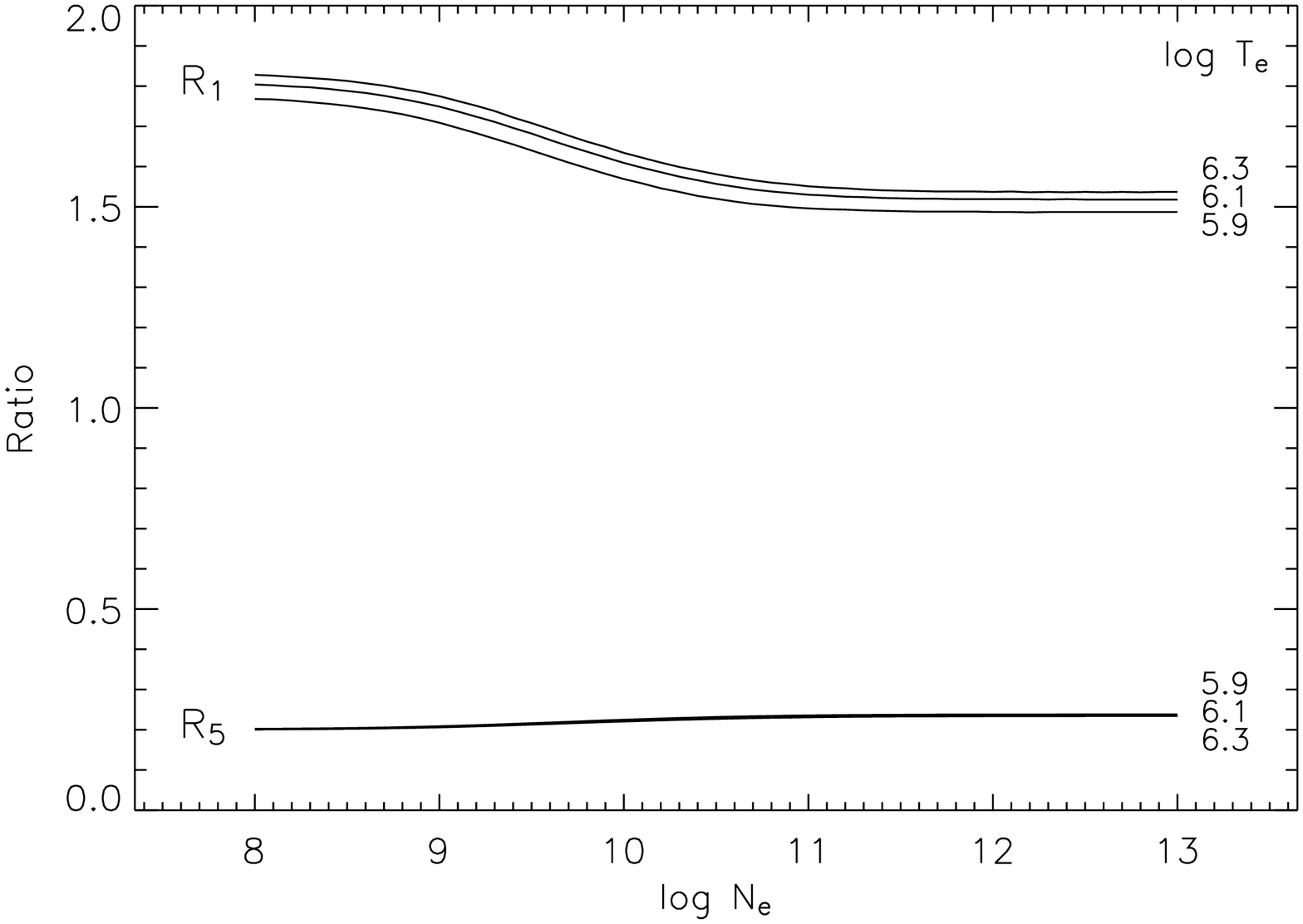}
\caption{
The theoretical \ion{Fe}{11}
emission line intensity ratios
R$_{1}$ = I(180.38 \AA)/I(188.21 \AA) and R$_{5}$ = I(188.30 \AA)/I(188.21 \AA),
where I is in energy units,
plotted as a function of logarithmic electron density
(N$_{e}$ in cm$^{-3}$) at the temperature
of maximum \ion{Fe}{11} fractional abundance in ionization
equilibrium, T$_{e}$ = 10$^{6.1}$ K (Mazzotta et al.
1998), plus $\pm$0.2 dex about this
value. 
}
\end{figure}

\begin{figure}
\includegraphics[scale=0.3,angle=90]{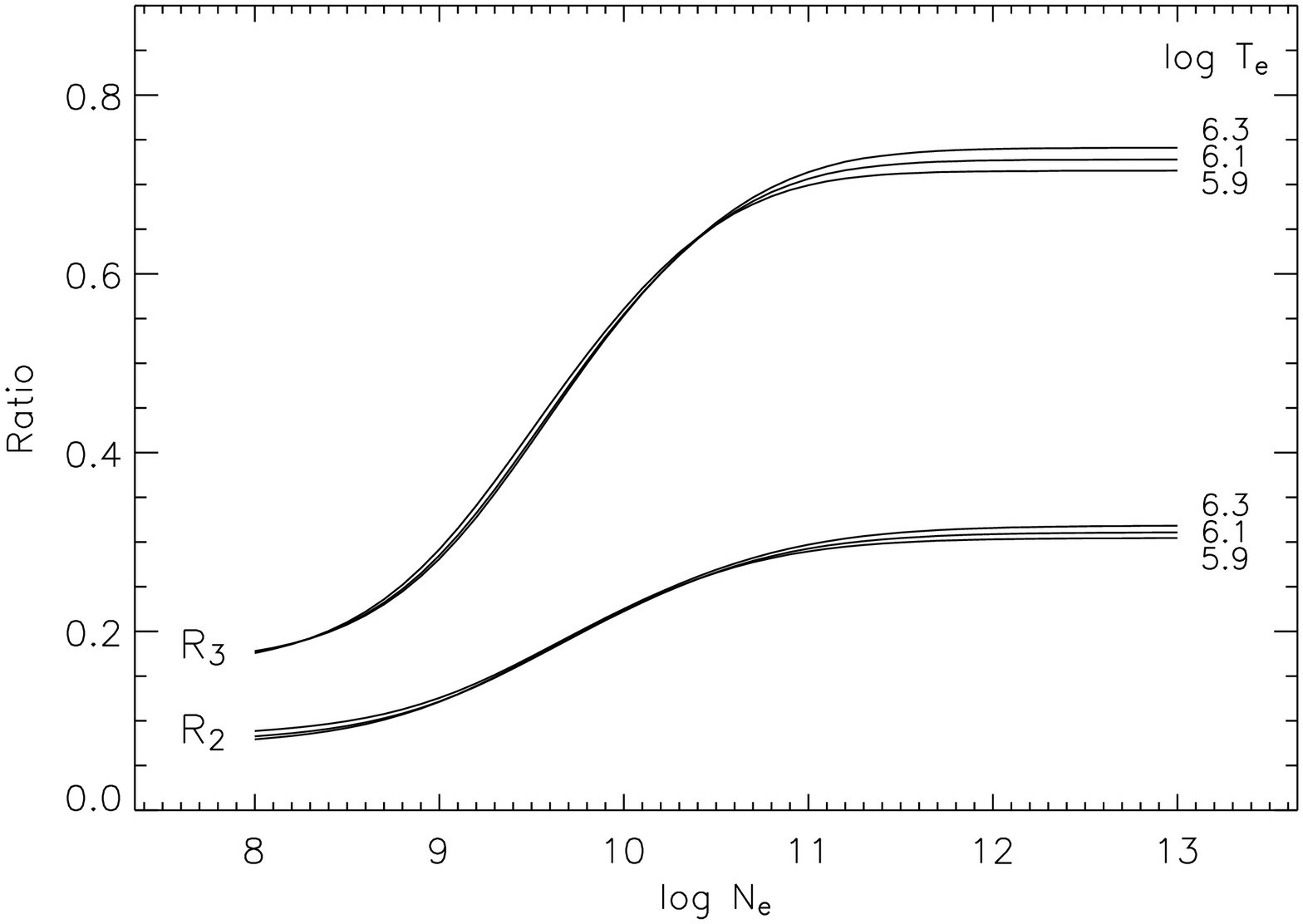}
\caption{
The theoretical \ion{Fe}{11}
emission line intensity ratios
R$_{2}$ = I(181.13 \AA)/I(188.21 \AA) and R$_{3}$ = I(182.17 \AA)/I(188.21 \AA),
where I is in energy units,
plotted as a function of logarithmic electron density
(N$_{e}$ in cm$^{-3}$) at the temperature
of maximum \ion{Fe}{11} fractional abundance in ionization
equilibrium, T$_{e}$ = 10$^{6.1}$ K (Mazzotta et al.
1998), plus $\pm$0.2 dex about this
value. 
}
\end{figure}

\begin{figure}
\includegraphics[scale=0.3,angle=90]{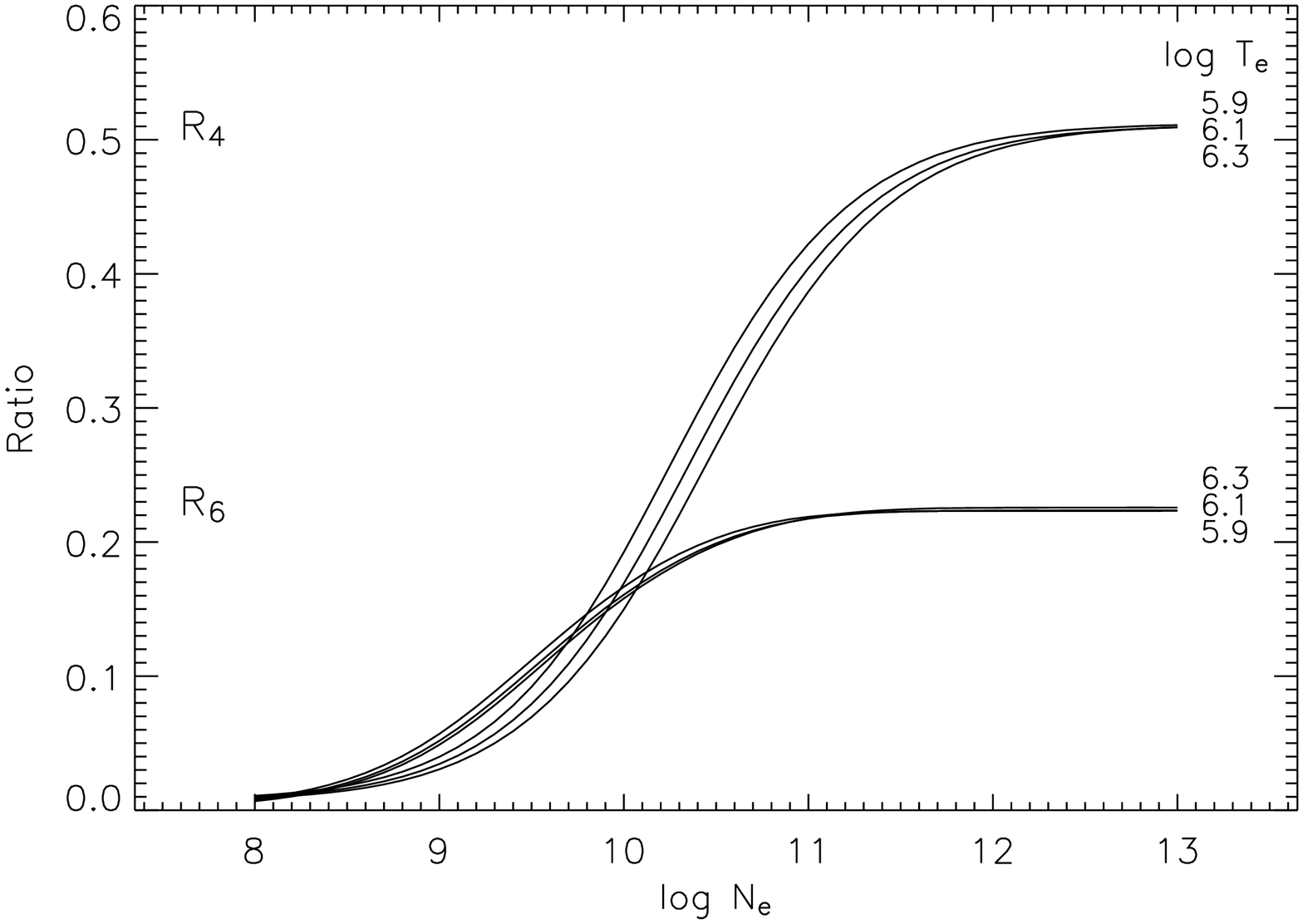}
\caption{
The theoretical \ion{Fe}{11}
emission line intensity ratios
R$_{4}$ = I(184.80 \AA)/I(188.21 \AA) and R$_{6}$ = I(189.00 \AA)/I(188.21 \AA),
where I is in energy units,
plotted as a function of logarithmic electron density
(N$_{e}$ in cm$^{-3}$) at the temperature
of maximum \ion{Fe}{11} fractional abundance in ionization
equilibrium, T$_{e}$ = 10$^{6.1}$ K (Mazzotta et al.
1998), plus $\pm$0.2 dex about this
value. 
}
\end{figure}

\begin{figure}
\includegraphics[scale=0.3,angle=90]{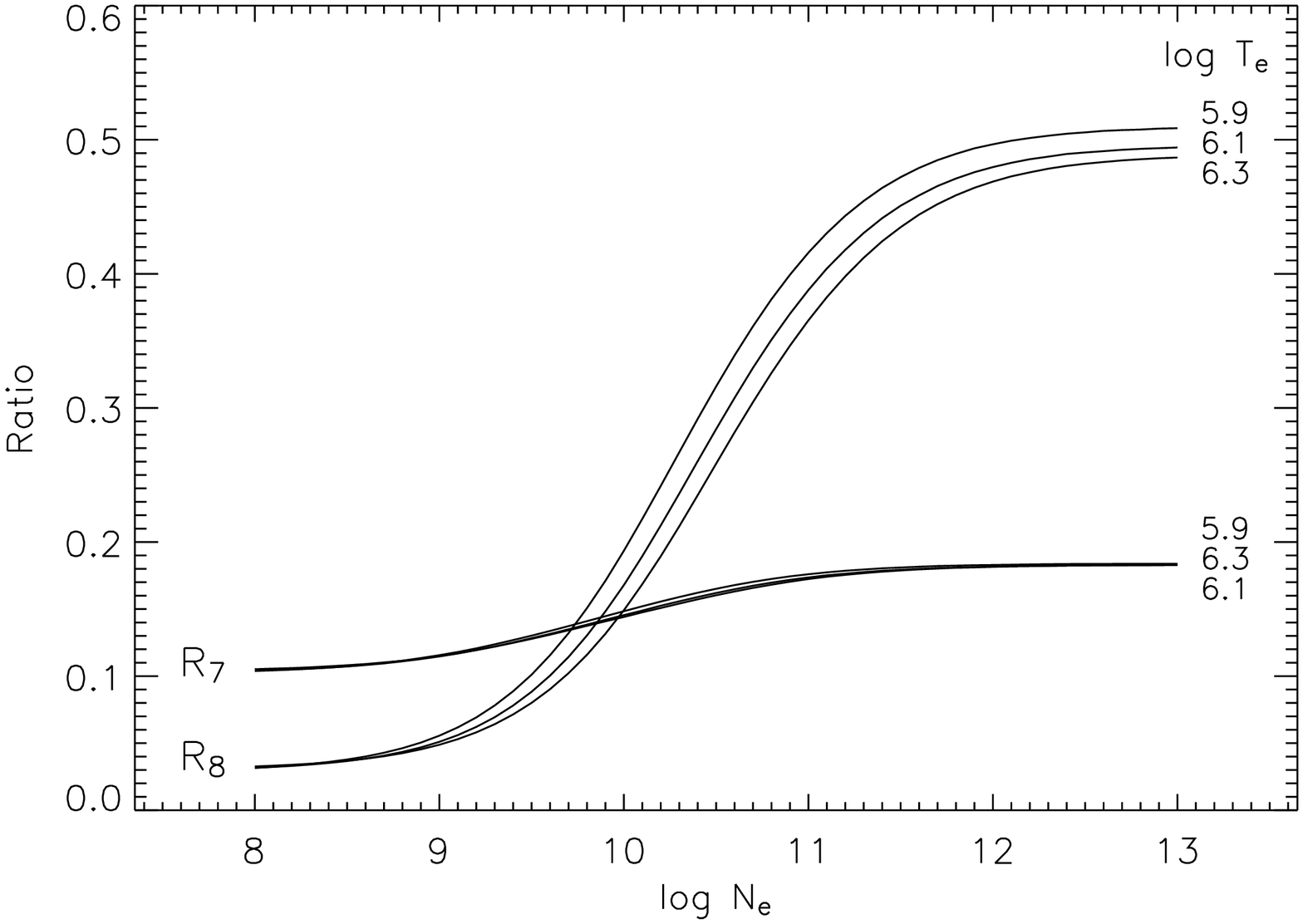}
\caption{
The theoretical \ion{Fe}{11}
emission line intensity ratios
R$_{7}$ = I(189.19 \AA)/I(188.21 \AA) and R$_{8}$ = I(202.42 \AA)/I(188.21 \AA),
where I is in energy units,
plotted as a function of logarithmic electron density
(N$_{e}$ in cm$^{-3}$) at the temperature
of maximum \ion{Fe}{11} fractional abundance in ionization
equilibrium, T$_{e}$ = 10$^{6.1}$ K (Mazzotta et al.
1998), plus $\pm$0.2 dex about this
value. 
}
\end{figure}

\begin{figure}
\includegraphics[scale=0.3,angle=90]{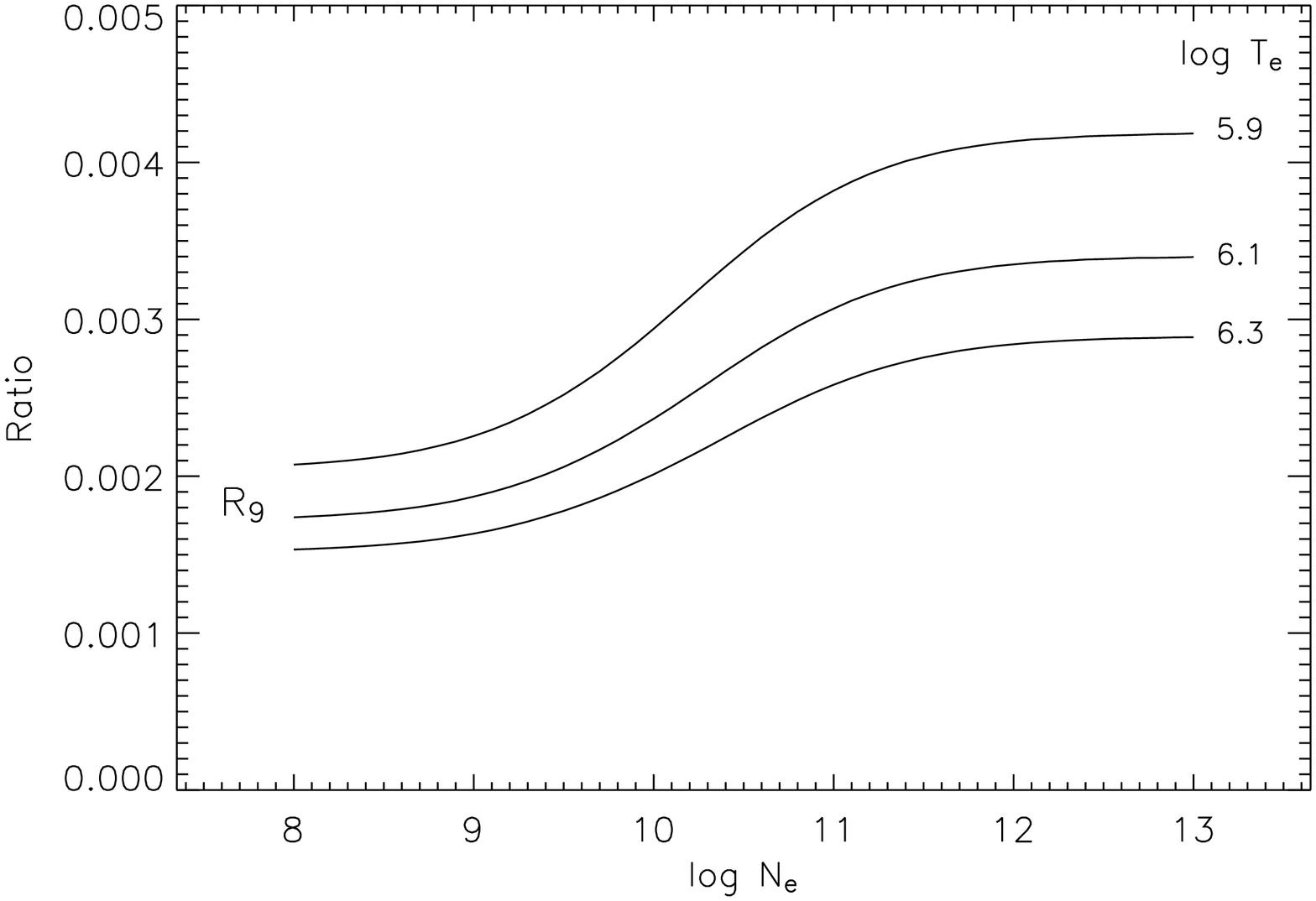}
\caption{
The theoretical \ion{Fe}{11}
emission line intensity ratio
R$_{9}$ = I(223.00 \AA)/I(188.21 \AA),
where I is in energy units,
plotted as a function of logarithmic electron density
(N$_{e}$ in cm$^{-3}$) at the temperature
of maximum \ion{Fe}{11} fractional abundance in ionization
equilibrium, T$_{e}$ = 10$^{6.1}$ K (Mazzotta et al.
1998), plus $\pm$0.2 dex about this
value. 
}
\end{figure}


\clearpage


\begin{deluxetable}{lcl}
\tablecaption{\ion{Fe}{11} line identifications in the SERTS 1995 active 
region spectrum}
\tablehead{
\colhead{Wavelength (\AA)} &
 \colhead{Transition} &
 \colhead{Note\tablenotemark{a}}
}                                                
\startdata
180.38 & 3{\em s}$^{2}$3{\em p}$^{4}$ $^{3}$P$_{2}$--3{\em s}$^{2}$3{\em 
p}$^{3}$($^{4}$S)3{\em d} $^{3}$D$_{3}$ & Blended with first-order \ion{Fe}{16} line
\\
181.13 & 3{\em s}$^{2}$3{\em p}$^{4}$ $^{3}$P$_{0}$--3{\em s}$^{2}$3{\em 
p}$^{3}$($^{4}$S)3{\em d} $^{3}$D$_{1}$ & \nodata
\\
182.17 & 3{\em s}$^{2}$3{\em p}$^{4}$ $^{3}$P$_{1}$--3{\em s}$^{2}$3{\em 
p}$^{3}$($^{4}$S)3{\em d} $^{3}$D$_{2}$ & \nodata \\
184.80 & 3{\em s}$^{2}$3{\em p}$^{4}$ $^{1}$D$_{2}$--3{\em s}$^{2}$3{\em 
p}$^{3}$($^{2}$D)3{\em d} $^{1}$D$_{2}$ & \nodata \\
188.21 & 3{\em s}$^{2}$3{\em p}$^{4}$ $^{3}$P$_{2}$--3{\em s}$^{2}$3{\em 
p}$^{3}$($^{2}$D)3{\em d} $^{3}$P$_{2}$ & \nodata \\
188.30 & 3{\em s}$^{2}$3{\em p}$^{4}$ $^{3}$P$_{2}$--3{\em s}$^{2}$3{\em 
p}$^{3}$($^{2}$D)3{\em d} $^{3}$S$_{1}$ & Tentative identification as \ion{Fe}{11} \\
189.00 & 3{\em s}$^{2}$3{\em p}$^{4}$ $^{3}$P$_{1}$--3{\em s}$^{2}$3{\em 
p}$^{3}$($^{2}$D)3{\em d} $^{3}$P$_{0}$ & Unidentified \\
189.19 & 3{\em s}$^{2}$3{\em p}$^{4}$ $^{3}$P$_{1}$--3{\em s}$^{2}$3{\em 
p}$^{3}$($^{2}$D)3{\em d} $^{3}$P$_{1}$ & Tentative identification as \ion{Fe}{11} \\
189.72 & 3{\em s}$^{2}$3{\em p}$^{4}$ $^{3}$P$_{0}$--3{\em s}$^{2}$3{\em 
p}$^{3}$($^{2}$D)3{\em d} $^{3}$P$_{1}$ & \nodata \\
192.81 & 3{\em s}$^{2}$3{\em p}$^{4}$ $^{3}$P$_{1}$--3{\em s}$^{2}$3{\em 
p}$^{3}$($^{2}$D)3{\em d} $^{3}$P$_{2}$ & \nodata \\
192.88 & 3{\em s}$^{2}$3{\em p}$^{4}$ $^{3}$P$_{1}$--3{\em s}$^{2}$3{\em 
p}$^{3}$($^{2}$D)3{\em d} $^{3}$S$_{1}$ & Listed as \ion{Ca}{17} \\
193.51 & 3{\em s}$^{2}$3{\em p}$^{4}$ $^{3}$P$_{0}$--3{\em s}$^{2}$3{\em 
p}$^{3}$($^{2}$D)3{\em d} $^{3}$S$_{1}$ & Listed as \ion{Fe}{12} \\
198.56 & 3{\em s}$^{2}$3{\em p}$^{4}$ $^{1}$D$_{2}$--3{\em s}$^{2}$3{\em 
p}$^{3}$($^{2}$D)3{\em d} $^{3}$P$_{1}$ & Listed as \ion{S}{8}/\ion{Fe}{12} blend 
\\
202.42 & 3{\em s}$^{2}$3{\em p}$^{4}$ $^{1}$D$_{2}$--3{\em s}$^{2}$3{\em 
p}$^{3}$($^{2}$D)3{\em d} $^{1}$P$_{1}$ & Listed as \ion{Fe}{13} \\
223.00 & 3{\em s}$^{2}$3{\em p}$^{4}$ $^{1}$D$_{2}$--3{\em s}$^{2}$3{\em 
p}$^{3}$($^{2}$P)3{\em d} $^{3}$D$_{1}$ & Listed as \ion{Fe}{11}/\ion{Ca}{17}
blend
\\
\enddata
\tablenotetext{a}{From Brosius et al. (1998).}
\end{deluxetable}

\clearpage


\begin{deluxetable}{lcc}
\tablecaption{\ion{Fe}{11} line intensities and widths in the SERTS 1995 active 
region spectrum}
\tablehead{
\colhead{Wavelength (\AA)} &
 \colhead{Intensity\tablenotemark{a}} &
 \colhead{Line width (\AA)}
}                                                
\startdata
180.38 & 3474 $\pm$ 384.5 & 0.051 $\pm$ 0.003
\\
181.13 & 123.8 $\pm$ 35.3 & 0.054 $\pm$ 0.011
\\
182.17 & 231.5 $\pm$ 31.6 & 0.038 $\pm$ 0.004
\\
184.80 & 32.9 $\pm$ 11.8 & 0.029 $\pm$ 0.009
\\
188.21 & 369.1 $\pm$ 68.5 & 0.043 $\pm$ 0.006
\\
188.30 & 245.9 $\pm$ 44.8 & 0.048 $\pm$ 0.007
\\
189.00 & 30.8 $\pm$ 8.0 & 0.023 $\pm$ 0.004
\\
189.19 & 31.6 $\pm$ 8.5 & 0.032 $\pm$ 0.006
\\
189.72 & 25.9 $\pm$ 6.3 & 0.028 $\pm$ 0.005
\\
192.81 & 68.4 $\pm$ 9.5 & 0.043 $\pm$ 0.004
\\
192.88 & 38.5 $\pm$ 10.0 & 0.072 $\pm$ 0.014
\\
193.51 & 478.8 $\pm$ 54.6 & 0.049 $\pm$ 0.003
\\
198.56 & 52.0 $\pm$ 10.2 & 0.036 $\pm$ 0.005
\\
202.42 & 71.2 $\pm$ 16.8 & 0.032 $\pm$ 0.005
\\
223.00 & 78.2 $\pm$ 16.3 & 0.029 $\pm$ 0.004
\\
\enddata
\tablenotetext{a}{In erg cm$^{-2}$ s$^{-1}$ sr$^{-1}$.}
\end{deluxetable}

\clearpage


\begin{deluxetable}{lcc}
\tablecaption{\ion{Fe}{11} line ratios in the SERTS 1995 active 
region spectrum}
\tablehead{
\colhead{Line ratio} &
 \colhead{Observed} &
 \colhead{Theoretical\tablenotemark{a}}
}                                                
\startdata
R$_{1}$ = I(180.38 \AA)/I(188.21 \AA) & 9.4 $\pm$ 2.0 & 1.7 $\pm$ 0.3
\\
R$_{2}$ = I(181.13 \AA)/I(188.21 \AA) & 0.34 $\pm$ 0.12 & 0.16 $\pm$ 0.03
\\
R$_{3}$ = I(182.17 \AA)/I(188.21 \AA) & 0.63 $\pm$ 0.15 & 0.39 $\pm$ 0.08
\\
R$_{4}$ = I(184.80 \AA)/I(188.21 \AA) & 0.089 $\pm$ 0.036 & 0.067 $\pm$ 0.013
\\
R$_{5}$ = I(188.30 \AA)/I(188.21 \AA) & 0.67 $\pm$ 0.17 & 0.21 $\pm$ 0.04
\\
R$_{6}$ = I(189.00 \AA)/I(188.21 \AA) & 0.083 $\pm$ 0.026 & 0.093 $\pm$ 0.019
\\
R$_{7}$ = I(189.19 \AA)/I(188.21 \AA) & 0.086 $\pm$ 0.028 & 0.13 $\pm$ 0.03
\\
R$_{10}$ = I(189.72 \AA)/I(188.21 \AA) & 0.070 $\pm$ 0.021 & 0.11 $\pm$ 0.02
\\
R$_{14}$ = I(192.81 \AA)/I(188.21 \AA) & 0.19 $\pm$ 0.04 & 0.19 $\pm$ 0.04
\\
R$_{11}$ = I(192.88 \AA)/I(188.21 \AA) & 0.10 $\pm$ 0.03 & 0.072 $\pm$ 0.014
\\
R$_{12}$ = I(193.51 \AA)/I(188.21 \AA) & 1.3 $\pm$ 0.3 & 0.020 $\pm$ 0.004
\\
R$_{13}$ = I(198.56 \AA)/I(188.21 \AA) & 0.14 $\pm$ 0.04 & 0.055 $\pm$ 0.011
\\
R$_{8}$ = I(202.42 \AA)/I(188.21 \AA) & 0.19 $\pm$ 0.06 & 0.078 $\pm$ 0.016
\\
R$_{9}$ = I(223.00 \AA)/I(188.21 \AA) & 0.21 $\pm$ 0.06 & 0.0020 $\pm$ 0.0004
\\
\enddata
\tablenotetext{a}{Determined from Figures 5--9 at N$_{e}$ = 10$^{9.4}$ cm$^{-3}$.}
\end{deluxetable}


\begin{thebibliography}{}

\bibitem[]{563} Aggarwal,  K. M., \& Keenan, F.P. 2003a, \mnras, 338, 412

\bibitem[]{565} Aggarwal, K. M., \& Keenan, F. P. 2003b, \aap, 399, 799

\bibitem[]{567} Behring, W. E., Cohen, L., \& Feldman, U. 1972,
\apj, 175, 493

\bibitem[]{570} Behring, W. E., Cohen, L., Feldman, U., \& Doschek, G. A. 1976,
\apj, 203, 521

\bibitem[]{573} Bhatia, A. K., \& Doschek, G. A. 1996, At. Data Nucl. Data Tables, 64, 183

\bibitem[]{575} Bhatia, A. K., Doschek, G. A., \& Eissner, W. 2002, At. Data Nucl. Data Tables, 
82, 211

\bibitem[]{578} Brosius, J. W., Davila, J. M., Thomas, R. J., \& Monsignori-Fossi, B. C.
1996, \apjs, 106, 143

\bibitem[]{581} Brosius, J. W., Davila, J. M., \& Thomas, R. J. 1998,
\apjs, 119, 255

\bibitem[]{584} Deb, N. C., \& Tayal, S. S. 1998, At. Data Nucl. Data Tables, 69, 161

\bibitem[]{586} Dere, K. P. 1978, \apj, 221, 1062

\bibitem[]{590} Dere, K. P., Landi, E., Mason, H. E., Monsignori-Fossi, B. C., 
\& Young, P. R. 1997, \aaps, 125, 149

\bibitem[]{593} Dufton, P. L. 1977, Comp. Phys. Commun., 13, 25

\bibitem[]{595} Dufton, P. L., Berrington, K. A., Burke, P. G., \& Kingston,
A. E. 1978, \aap, 62, 111

\bibitem[]{598} Dufton, P. L., Kingston, A. E., Doyle, J. G., \& Widing, K. G.
1983, \mnras, 205, 81

\bibitem[]{601} Gupta, G. P., \& Tayal, S. S. 1999a, \apj, 510, 1078

\bibitem[]{603} Gupta, G. P., \& Tayal, S. S. 1999b, \apjs, 123, 295

\bibitem[]{605} Howarth, I. D., Murray, J., \& Mills, D. 1994, Starlink User
Note No. 50.15

\bibitem[]{608} Jup\'{e}n, C., Isler, R. C., \& Tr\"{a}bert, E. 1993,
\mnras, 264, 627

\bibitem[]{611} Kastner, S. O., \& Mason, H. E. 1978, \aap, 67, 119

\bibitem[]{613} Keenan, F. P., Aggarwal, K. M., Katsiyannis, A. C., 
\& Reid, R. H. G. 2003, \solphys, 217, 225

\bibitem[]{616} Keenan, F. P., Botha, G. J. J., Matthews, A., Lawson, K. D., 
\& Coffey, I. H. 2000, \mnras, 318, 37

\bibitem[]{619} Kelly, R. L. 1987, J. Phys. Chem. Ref. Data, 16, Suppl. 1

\bibitem[]{621} Landman, D. A. 1980, \apj, 240, 709

\bibitem[]{623} Mazzotta, P., Mazzitelli, G., Colafrancesco, S.,
\& Vittorio, N. 1998, \aaps, 133, 403

\bibitem[]{626} Neupert, W. M., Epstein, G. L., Thomas, R. J., \&
Thompson, W. T. 1992, \solphys, 137, 87

\bibitem[]{629} Shirai, T., Funatake, Y., Mori, K., Sugar, J., Wiese, W. L.,
\& Nakai, Y. 1990, J. Phys. Chem. Ref. Data, 19, 127

\bibitem[]{632} Thomas, R. J., \& Neupert, W. M. 1994, \apjs, 91, 461

\bibitem[]{634} Young, P. R., Del Zanna, G., Landi, E., Dere, K. P., Mason, H. E.,
\& Landini, M. 2003, \apjs, 144, 135

\end{thebibliography}
\end{document}